\documentclass[10pt,review]{elsarticle}

\usepackage{lineno,hyperref}
\modulolinenumbers[5]

\journal{Journal of Safety Science}

\usepackage{booktabs}

\usepackage{tikz}
\usetikzlibrary{positioning, shapes.geometric, arrows.meta, fit, backgrounds, calc, decorations.pathreplacing}

\usepackage{comment}

\usepackage{algorithm}
\usepackage[noend]{algpseudocode}

\usepackage{amsmath}
\usepackage{amssymb}

\usepackage{amsfonts}

\usepackage{makecell}

\usepackage{xcolor}

\usepackage{appendix}

\usepackage{amsmath,amssymb}

\usepackage[sfdefault]{atkinson} %% Option 'sfdefault' if the base

\begin{document}

\begin{frontmatter}

\title{Is Your Model Risk ALARP? Evaluating Prospective Safety-Critical Applications of Complex Models}

\author[address_1,address_2,address_3]{Domenic Di Francesco}
%\ead[url]{ddifrancesco"turing.ac.uk}

\author[address_4]{Alan Forest}
\author[address_3]{Fiona McGarry}
\author[address_3]{Nicholas Hall}
\author[address_1,address_5]{Adam Sobey}


\address[address_1]{The Alan Turing Institute for Artificial Intelligence and Data Science, The British Library, 2QR, John Dodson House, 96 Euston Rd, London NW1 2DB}

\address[address_2]{Department of Civil Engineering, Cambridge University, Trumpington Street, CB2 1PZ}

\address[address_3]{Health and Safety Executive, Harpur Hill, Buxton SK17 9JN}

\address[address_4]{Credit Research Centre, University of Edinburgh Business School, 29 Buccleuch Place, Edinburgh, EH8 9JS}

\address[address_5]{Department of Engineering, Southampton University, University Rd, Southampton SO17 1BJ}

\begin{abstract}
The increasing availability of advanced computational modelling offers new opportunities to improve safety, efficacy, and emissions reductions. Application of complex models to support engineering decisions has been slow in comparison to other sectors, reflecting the higher consequence of unsafe applications. 

Adopting a complex model introduces a model risk, namely the expected consequence of incorrect or otherwise unhelpful outputs. This should be weighed against the prospective benefits that the more sophisticated model can provide, also accounting for the non-zero risk of existing practice. Demonstrating when the model risk of a proposed machine learning application is As Low As Reasonably Practicable (ALARP) can help ensure that safety-critical industries benefit from complex models where appropriate while avoiding their misuse. An example of automated weld radiograph classification is presented to demonstrate how this can be achieved by combining statistical decision analysis, uncertainty quantification, and value of information.

\end{abstract}

\begin{keyword}
AI Safety, Decision Analysis Under Uncertainty, Asset Management, Model Risk.

%\texttt{elsarticle.cls}\sep \LaTeX\sep Elsevier \sep template
%\MSC[2010] 00-01\sep  99-00

\end{keyword}

\end{frontmatter}

%\linenumbers

\begin{table} % Added [htbp] for better float placement
    \centering
    \label{tab:nomenclature}
    \begin{tabular}{c|c}
        Symbol & Meaning \\
        \hline
        % --- Greek Letters ---
        $\alpha$  & Dirichlet distribution parameters \\
        $\eta$ & learning rate \\ 
        $\theta$ & model reliability parameters  \\
        $\pi_i$ & mixture model component weights \\
        
        % --- Mathematical Symbols ---
        $\mathcal{C}[d(m_o),s]$ & cost associated with decision $d(m_o)$ in scenario $s$\\
        $\mathcal{C}_{Prior}$ & prior expected cost \\
        $\mathcal{C}_{Pre-posterior}$ & pre-posterior expected cost \\
        $\mathcal{C}_{fail}$ & failure cost \\
        $\mathcal{I}(s,m_o)$ & impact when model output $m_o$ occurs in scenario $s$ \\
        $\mathcal{L}(\cdot)$ & loss function  \\
        $R_m$ & risk associated with the use of model $m$ \\

        % --- English Alphabet Variables ---
        $C_{i,:}$ & row $i$ of confusion matrix \\
        $f_i$ & component distributions in mixture model \\
        $m$ & a proposed model from a set of available models, $M$  \\
        $m^*$ & model with the lowest model risk, $R_m$  \\
        $m_o$ & specific output from a model $m$ \\
        $M$ & set of available models\\
        $N_i$ & total count for true class $i$ from a classification model \\
        $s$ & a specific scenario or true state of the system\\
        $S$ & set of all possible scenarios  \\
        $S_c(z)$ & class score, $\Pr(y=c|z)$, for class $c$ given input $z$  \\
        $S_{i,j}$ & saliency value associated with input pixel $z_{i,j}$  \\
        $y_{cf}$ & desired counterfactual prediction  \\
        $z$ & input data to a model \\
        
    \end{tabular}
    \caption{Nomenclature}
\end{table}

\newpage

\section{Introduction and context} \label{sect:intro}

\subsection{Concepts and definitions} \label{sect:concepts}

Introducing complex models into safety-critical workflows presents a fundamental trade-off. An overly pessimistic approach prevents technological advances that could benefit society, while excessive optimism increases catastrophic failure risks, potentially causing environmental damage, injury, or loss of life. Quantitative risk assessment provides a framework for duty holders to evaluate whether proposed modelling interventions offer sufficient benefits to justify their associated risks.

For the purposes of this work, the following definitions apply:
\begin{itemize}
    \item \textbf{complex model}: A mathematical or computational representation of a true system. These may include Bayesian models of uncertainty, deep neural networks, ensemble methods, digital twins, or other advanced machine learning techniques whose internal operations are not as transparent or easily interpretable as established, standardised alternatives in engineering.
    \item \textbf{model risk}: The expected adverse impact from incorrect or misused model outputs. By incorporating downstream benefits, model risk enables quantitative comparison of alternative approaches.
    \item As Low As Reasonably Practicable (\textbf{ALARP}): A UK regulatory principle requiring risk reduction until further reduction becomes disproportionate to benefits gained. This establishes a threshold for \emph{acceptable risk} balancing safety with practical and economic considerations.
\end{itemize}

\subsection{Summary of relevant scientific literature} \label{sect:literature}

UK government guidance  has highlighted the need for balancing AI regulation to mitigate societal harms without impeding innovation \cite{DSIT2024}. Key principles from regulatory publications \cite{HMTreasury2015, UKGovernment2023} include:

\begin{itemize}
    \item \textbf{RIGOUR}: models can be assessed with respect to the extent to which they are Repeatable, Independent, Grounded in reality, Objective, Uncertainty-managed, Robust. The Aqua Book \cite{HMTreasury2015} in particular dedicates significant attention to identifying, quantifying, and communicating uncertainty.
    \item \textbf{proportionate quality assurance (QA)}: the principle that the level of scrutiny (of model inputs, methods and outputs) should be proportional to the risks involved.
    \item \textbf{defined roles and responsibilities}: risk identification, mitigation and assurance can benefit from input from multiple (both internal and independent) layers.
\end{itemize}

The financial industries have developed comprehensive guidance on model risk, following the global financial crash in 2008 The Bank of England’s review of the crisis \cite{FSA2009}, cited (amongst other factors) a misplaced trust in seemingly complex modelling strategies, that actually failed to adequately account for tail behaviour (extreme, low-probability events) and inter-dependencies in risk models:

%The 2008 financial crisis prompted comprehensive model risk guidance in financial sectors \cite{Letter11_7}. Regulatory review \cite{FSA2009} identified misplaced confidence in complex models that failed to capture tail behaviour (extreme, low-probability events) and interdependencies:

\begin{quote}
    \emph{Mathematical sophistication ended up not containing risk, but providing false assurance that other prima facie indicators  ...could be safely ignored.}
\end{quote}

In response to such findings, model risk is now elevated in banks as a principal risk and is managed by specific regulation in the US \cite{Letter11_7} and the UK \cite{BankofEngland2023}. This regulation, and similar professional standards, have caused a deep and lasting change in bank modelling culture internationally. The UK regulation identified five principles for managing model risk, and this wider environmental view of model risk is especially important when the models are automated, complex or black-box \cite{BankofEngland2022}.

\begin{itemize}
    \item \textbf{model identification and model risk classification}: all models in a bank should be inventoried, and their risk managed systematically.
    \item \textbf{governance}: models must be governed within a model risk framework. The board is accountable for the banks’ models and senior managers are responsible for the management of models risk.
    \item \textbf{model development, implementation and use}: model development must meet high standards for design, data quality, testing, and documentation.
    \item \textbf{independent model validation}: an independent validation function must provide ongoing, and effective challenge to model development and use. This includes performance monitoring.
    \item \textbf{model risk mitigants}: models must sit within a strict control environment of testing, monitoring, intervention, exception, use restrictions and escalation.
\end{itemize}

Beyond finance, model cards \cite{Mitchell_2019} have been proposed for comprehensive performance reporting, contrasting with tendencies to highlight only best-case scenarios \cite{Saidi2025}. Safety-critical domains require deeper understanding of failure modes. This work extends the model card concept through statistical analysis of model reliability, enabling proactive risk management.

High-level guidance exists in standards \cite{BSI2023} for documenting risk management processes, but regulatory (external) interventions require clarity and relevance to facilitate the development of (internal) constructive and sustainable practices \cite{Stahl2025}. These are considered to be missing in current guidance due to the absence of technical details. The Alan Turing Institute developed a platform for specifying functional requirements from complex models \cite{Institute2024}, demonstrated in healthcare applications \cite{burr_2024}. Such tools can help address the challenge of risk identification in novel applications such as model risk.

Explainable AI techniques help identify model risk sources. Counterfactual analysis generates adversarial examples \cite{Altmeyer2023}, with extensions incorporating interpretable concepts \cite{Abid2021} to identify human-understandable error patterns. Gradient-based methods produce saliency maps showing influential input regions, providing insight into opaque model workings \cite{kares2025}. These maps use standard backpropagation \cite{Simonyan14} or final convolutional layer gradients for noise reduction \cite{Selvaraju2019}. Such visual representations build trust and support risk mitigation in high-consequence applications, as demonstrated in Section \ref{sect:examples_raddt}.

The purpose of complex models is generally to support decision making. In this paper, the use of decision (influence) diagrams\footnote{which are an extension of Directed Acyclic Graphs (DAGs) to include available decisions/interventions and outcomes} is advocated for. Decision analysis is considered to require causal reasoning that can beyond the capability of complex models \cite{Felin}, which is considered particularly relevant for safety-critical engineering applications where decisions often require anticipating novel failure modes and making context-specific judgments that cannot be derived from historical data alone.

The analysis presented in this paper was completed using machine learning software library \emph{Flux} \cite{Innes2018}, in the \emph{Julia} programming language \cite{bezanson2017}, using \emph{Enzyme} for automatic differentiation \cite{Moses2020, Moses2021}. It focuses on the risk associated with model reliability, but a comprehensive review of model risk would also include a Failure Modes and Effects Analysis (FMEA) study \cite{Stavrou2025}.

\begin{comment}
In reinforcement learning, such as the asset management example in Section \ref{sect:examples_rl}, an agent explores the space of a decision making under uncertainty problem. Feedback from actions during training allow for a policy/strategy to be learned/approximated \cite{Sutton2020}. To address safety concerns of the application of, in particular black-box, policy approximations, a technique known as \emph{shielding} ensures defined unwanted behaviours are avoided. Despite these constraints, some research has indicated this method may even provide performance benefits in training \cite{Carr2023}.
\end{comment}

\section{Risk Management} \label{sect:risk}

\subsection{Introduction} \label{sect:risk_intro}

For the purposes of this paper, risk is defined as the expected consequence or impact of an activity, consistent with engineering applications in subsea \cite{DNV2021} and petrochemical \cite{AmericanPetroleumInstitute2008, AmericanPetroleumInstitute2016} industries. This quantification enables rank-ordering of decision alternatives \cite{DiFrancesco2025} and justifies interventions when projected benefits outweigh expected costs, conditional on the models employed. 

Engineering disciplines have established rigorous verification practices for structural and mechanical designs to demonstrate safety \cite{DNV2021a}. In this paper, the case is made that these same verification principles should extend to model risk assessment, ensuring consistent safety standards across all aspects of engineering systems.

\subsection{Model risk} \label{sect:risk_of_models}

%Model risk is influenced by our understanding of relevant performance metrics, governance structures, and implementation practices. While it can be mitigated through various strategies, a formal comparison between alternative approaches enables determination of which model is most suitable for a specific application. Safety regulators may request evidence that model risk has been considered and managed to be ALARP, applying the same principles used for more established engineering risks related to physical systems and degradation processes. This paper proposes a quantitative framework for assessing and managing model risk in safety-critical applications.

The risk associated with using model $m$ is denoted $\mathcal{R}_m$. As shown in Equation \ref{eq:risk}, model risk quantifies the expected impact $\mathcal{I}$ when the model is deployed across various scenarios $s \in S$. Following risk management conventions, desirable outcomes contribute negative values to minimise overall risk.

The risk formulation reflects the deployment context: scenarios occur with probability $\Pr(s)$, and within each scenario, the model produces output $m_o \in M_O$ with probability $\Pr(m_o \mid s)$. The product of these probabilities weights the impact of outcomes.

\begin{equation}
    \mathcal{R}_m = \int_{S} \int_{M_O} \Pr(s) \cdot \Pr(m_o \mid s) \cdot \mathcal{I}(s, m_o) \, dm_o \, ds
    \label{eq:risk}
\end{equation}

Where $\mathcal{I}(s, m_o)$ represents the impact when model output $m_o$ occurs in scenario $s$, effectively serving as a cost/utility function over the joint domain of scenarios and outputs. 

\begin{comment}
Some example considerations in assigning this value:
\begin{itemize}
    \item \textbf{quality of model output}: if a complex model performs better in a given scenario, this may result in new insights, better resource allocation, and more timely interventions. In some cases, the use of a complex model may facilitate a new type of analysis, that are beyond the scope of alternatives, resulting in new opportunities.
    \item \textbf{explainability}: how difficult is the misuse of various models to identify and correct? For a model with black box components, model error may not be evident until after subsequent action has been taken.
    \item \textbf{implementation}: what costs are associated with putting various models into production? These may include the costs of procuring data, computational resources for training, and monitoring costs.
\end{itemize}
\end{comment}

Model risk enables comparison between alternatives, for instance a standardised simple approach versus a complex black-box model. The risk-optimal model $m^*$ minimises the expected risk, see Equation \ref{eq:risk_opt_model}.

\begin{equation}
    m^{*} =  \mathop{\arg \min}\limits_{m \in M} \; \mathcal{R}_{m}
    \label{eq:risk_opt_model}
\end{equation}

Model risk evolves over time. Data distributions may drift due to changing operational environments, regulatory shifts, or evolution in the underlying system \cite{Lu2019}. Such drift can degrade model performance and invalidate initial risk assessments. Monitoring is therefore essential to detect changes and reassess the risk landscape. As drift occurs or models are retrained, the risk-based ranking of models may shift, potentially identifying different optimal solutions over time.

\section{Example: Automated evaluation of weld radiographs}\label{sect:examples_raddt}

\subsection{Introduction}

Welded steel represents a vast range of critical infrastructure globally, including bridges, buildings, pipelines, pressure vessels, and offshore platforms. The presence of stress concentrations (due to geometric misalignments), and tensile residual stresses mean that welding imperfections can become initiation sites for defects with the potential to result in catastrophic failures. Consequently, in safety critical industries, welds are generally inspected using various technologies to detect and size weld anomalies, including radiographs.

Radiographic testing is often specified as a requirement for some proportion of welds in a project. This is between $1\%$ and $100\%$ for tanks (depending on weld type, thickness, and material yield strength) \cite{BritishStandardsInstitute2004}. When anomalies are identified in a random sample of radiographs of welds in pipework, further testing is triggered  \cite{BritishStandardsInstitution1987}. Requirements for verification of qualification (competency) of testing personnel are specified in guidance for aerospace applications \cite{NASA2008}, \cite{NASA2020}. Acceptance rates generally vary with the type of anomaly, for example both cracking and lack of fusion are never permitted in pressure vessels, whereas porosity can be permissible depending on it's dimensions \cite{BritishStandardsInstitute2021}.

For a large construction project, it may take an inspector minutes to carefully review each image and locate and classify critical anomalies. Computer vision models can operate in near real-time, and are now considered to be an established technology in machine learning, and one potential application of this is to automate the classification of weld radiograph data. This would address the potential limitations of manual interpretation, such as inconsistency, speed, and effects of physiological fatigue, for a sufficiently reliable model.

\subsection{Problem description}

A dataset of radiographs was obtained from a previous research project \cite{Totino2023}, which demonstrated the feasibility of image classification models for this task \cite{Perri2023}. The question of whether a machine learning model \emph{can} classify weld radiographs is therefore considered to have been answered. The challenge considered here is understanding how to evaluate when it is a risk optimal solution in practice.

\begin{figure}
    \centering
    \includegraphics[width = \textwidth]{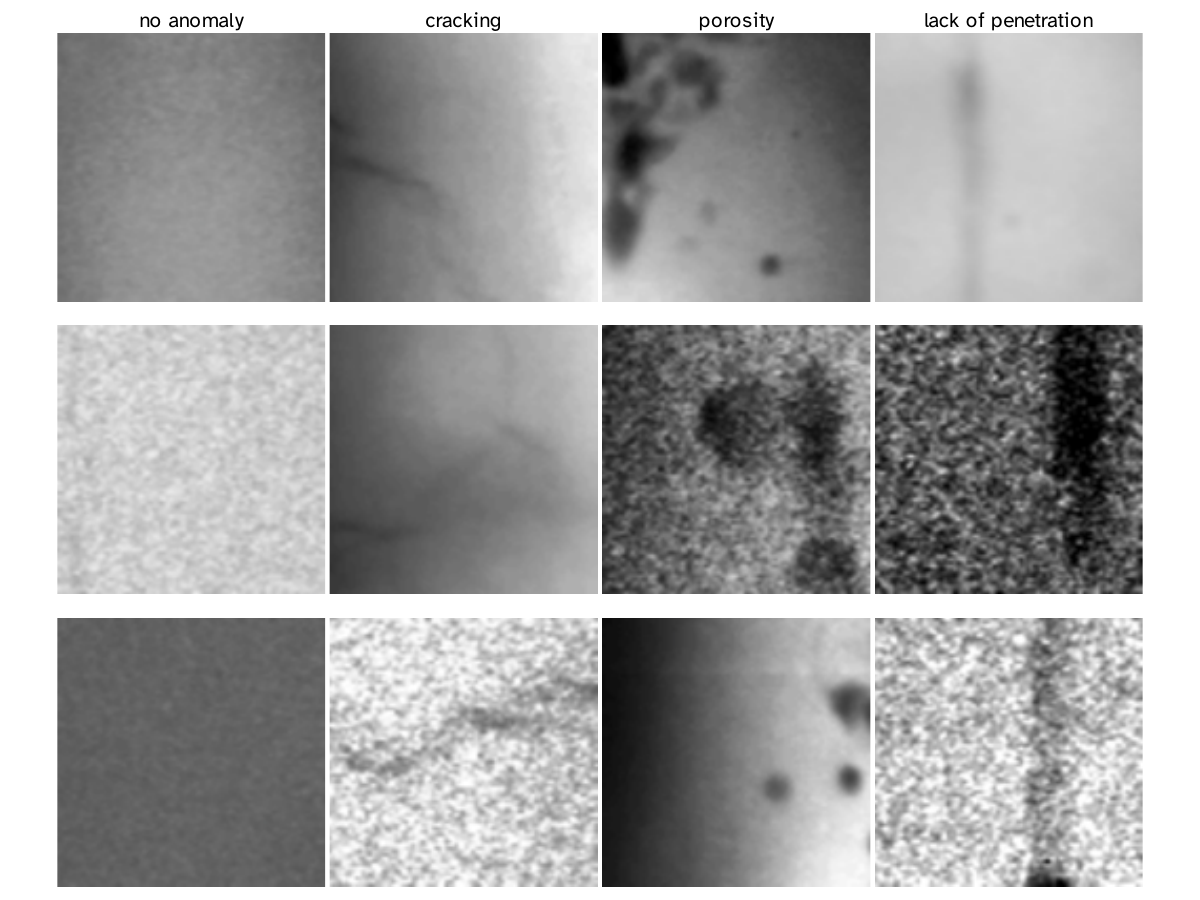}
    \caption{Example images of various types from the weld radiograph dataset}
    \label{fig:weld_examples}
\end{figure}

\begin{figure}
    \centering
    \includegraphics[width = \textwidth]{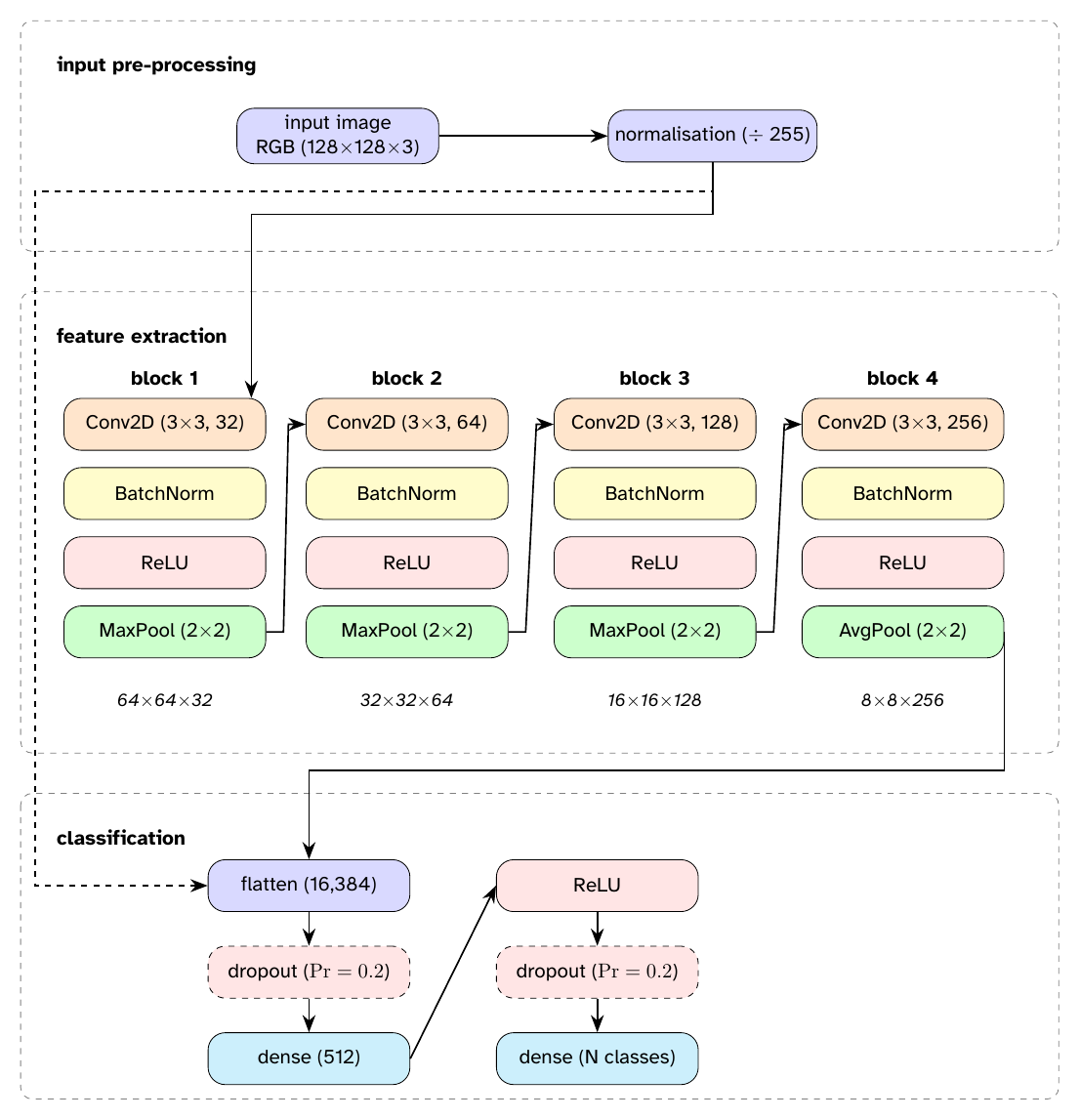}
    \caption{Schematic diagram of weld classification model structure}
    \label{fig:CNN_architecture}
\end{figure}

Weld radiographs with various types of damage (and in a nominally undamaged state) are used to train a classification model. A few examples of each type are shown in Figure \ref{fig:weld_examples}. The schematic structure of a proposed Convolutional Neural Network (CNN) image classification model is shown in Figure \ref{fig:CNN_architecture}. Key details of this model include: some pre-processing (input normalisation), a progressive filter depth to help identify both coarse and more subtle anomalies, and dropout to help prevent overfitting by creating an implicit ensemble effect.

\begin{comment}

\begin{itemize}
    \item \textbf{pre-processing}. Input normalisation re-scales pixel values from $[0,255]$ to $[0,1]$. This improves numerical stability when computing gradients during training and establishes a consistent input scale.
    
    \item \textbf{progressive filter depth $(32 \rightarrow 64 \rightarrow 128 \rightarrow 256)$}. The architecture uses four sequential convolutional blocks with increasing filter counts. Each filter acts like a specialised detector - early layers with fewer filters (32) identify basic patterns like edges and transitions, while later layers with more filters extract increasingly complex features. This hierarchical detection capability allows the network to distinguish between various weld defect types, from coarse lack of penetration defects to fine instances of porosity.

    \item \textbf{batch normalisation}. This technique standardises the inputs to activation functions across each batch of images. This improves training stability, particularly for input value distributions that would otherwise be more dispersed.
    
    \item \textbf{dropout}. The network transitions from spatial representation ($16,384$ neurons) to classification output ($N = 4$ damage states) via fully-connected layers. During training, dropout temporarily deactivates a sample of $20\%$ of neurons, creating an implicit ensemble effect that prevents overfitting. 
    
\end{itemize}

\end{comment}

\begin{figure}
    \centering
    \includegraphics[width = \textwidth]{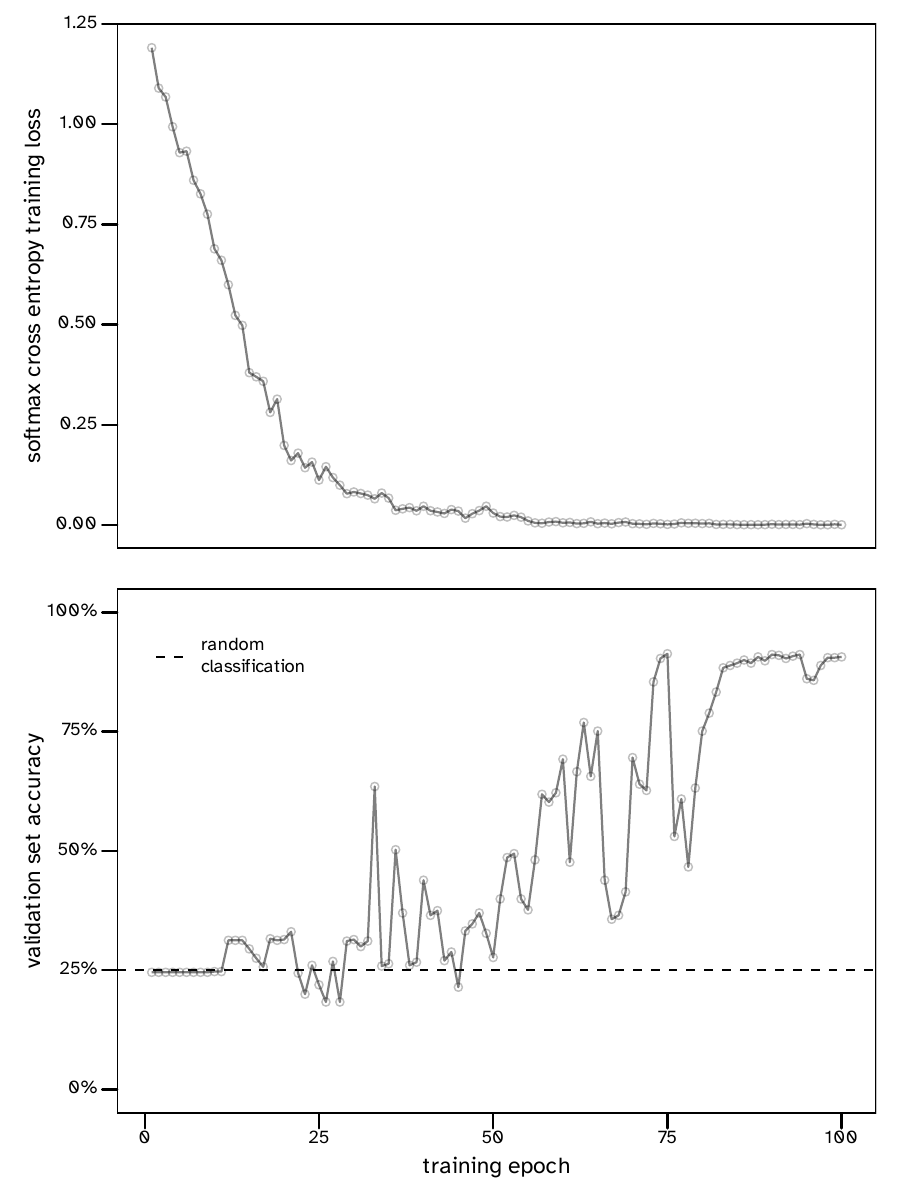}
    \caption{Weld classification model during training}
    \label{fig:CNN_training}
\end{figure}

\subsection{Model risk quantification} \label{sect:risk_quant}

\subsubsection{Decision Analysis} \label{sect:decisions}

Model risk arises only when models inform decisions. Even an unreliable model poses no risk beyond development costs if its outputs remain unused. Therefore, evaluating model risk requires identifying model-decision boundaries.

\begin{figure}
    \centering
    \includegraphics[width=\textwidth]{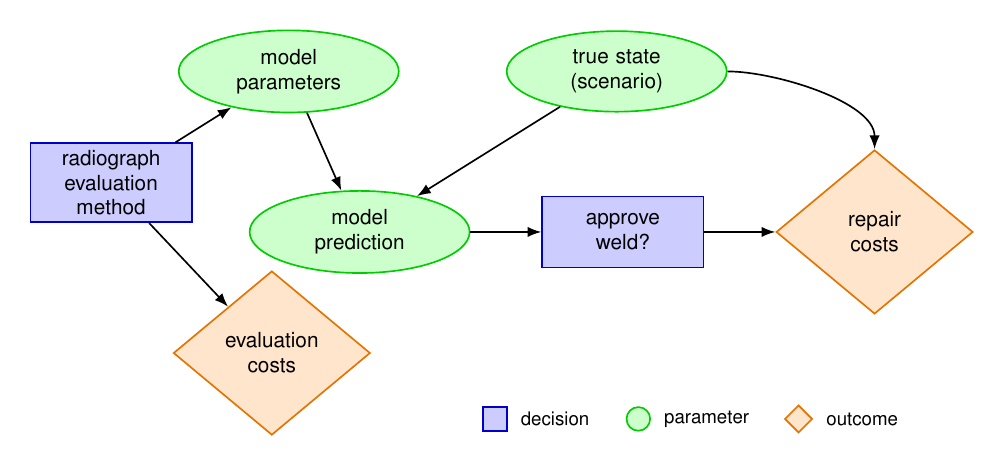}
    \caption{Influence diagram describing decision problem of evaluation of radiographs for quality assurance of welds}
    \label{fig:inf_diag}
\end{figure}

Consider a quality assurance example where a complex model that has been trained to classify damage in weld radiographs is available. Figure \ref{fig:inf_diag} presents the decision problem as a decision diagram, with circular nodes representing uncertain parameters, rectangular nodes representing decisions, and diamond nodes representing costs.

For each scenario $s$ (the true damage category), the expected cost weights outcomes by their probabilities, see Equation \ref{eq:exp_cost_scenario}.

\begin{equation}
    \mathcal{R}_{m}(s) = \sum_{m_{o} \in M_{o}} \left( \underbrace{\mathcal{C} \left[ d(m_{o}), s \right]}_{\text{cost associated with output }} \times \underbrace{\Pr(m_{o} \mid s)}_{\text{probability of output}} \right)
    \label{eq:exp_cost_scenario}
\end{equation}

This follows from Equation \ref{eq:risk}, as a function of the true scenario, $s$, for a discrete number of possible model outputs, $m_{o}$, and financial costs being the impact considered. Here, $d(m_o)$ represents the decision rule triggered by a given model output. When these costs encompass all decision-relevant outputs, this expression quantifies scenario-specific model risk.

This analysis compares three models (detailed below) using the cost structure in Table \ref{tab:fundamental_costs}. 

\begin{enumerate}
    \item \textbf{manual (perfect) evaluation}: Here the weld is assessed by a suitably trained inspector. The reliability of this approach will vary based on the extent of the damage (classification difficulty) and various human factors. However, in this initial example, it is assumed that an inspector classifies weld anomalies without error from radiographs. Consequently, the predicted damage state will always match the true state, and if an anomaly is identified, it is scheduled for the appropriate repair.
    \item \textbf{fully automated evaluation}: using samples from the posterior distribution of $\Pr(m_{o} \mid s)$ for each scenario, calculate the corresponding costs of scheduling only predicted anomalies for repair. If, during the repair, a different type of damage is identified, the costs of both repairs is incurred.
    \item \textbf{a hybrid approach}: in instances where the classification model predicts more consequential damage (cracking or a lack of penetration), then the weld is sent to a (perfect) manual inspector before a decision is made.
\end{enumerate}

In practice, the reliability of manual inspections will vary based on the extent of the damage, human error probabilities, and performance shaping factors. In this example, a perfect performance assumption (though unrealistic) represents the most challenging benchmark for evaluating the complex model. It is important to note that this analysis is equally compatible with alternative input models.

\begin{table}[htbp]
    \centering
    \begin{tabular}{lr}
    \toprule
    \textbf{Activity description} & \textbf{Cost (\pounds)} \\
    \midrule
    Manual evaluation of one radiograph & 350 \\
    \addlinespace
    Repairing cracking & 1000 \\
    Repairing lack of penetration & 3000 \\
    Repairing porosity & 500 \\
    \addlinespace
    Failure due to unrepaired lack of penetration & $\mathcal{C}_{\text{fail}}$ \\  % C_fail
    Failure due to unrepaired cracking & $1/2 \times \mathcal{C}_{\text{fail}}$ \\ % C_fail / 2
    Failure due to unrepaired porosity & $1/10 \times \mathcal{C}_{\text{fail}}$ \\   % C_fail / 10
    % Note: Assuming no separate cost for running the automated model itself, or it's negligible.
    % If there were a 'cost of approving a correct NA weld', it's 0 as per your C_model.
    \bottomrule
    \end{tabular}
    \caption{Inputs for weld radiograph decision analysis. These costs are combined differently depending on the chosen analysis strategy (manual, model-based, or hybrid) and the specific true state versus predicted state scenario}
    \label{tab:fundamental_costs}
\end{table}

The failure cost $\mathcal{C}_{\text{fail}}$ is characterised using a mixture model to account for uncertainty in consequences, see Equations \ref{eq:c_fail_general}, \ref{eq:c_fail_mixing_weights}, and \ref{eq:c_fail_components}.

\begin{equation}
    \mathcal{C}_{\text{fail}} \sim \sum_{i=1}^{2} \pi_i \cdot f_i
    \label{eq:c_fail_general}
\end{equation}

\begin{equation}
    (\pi_1, \pi_2) \sim \text{Dirichlet}(9, 3)
    \label{eq:c_fail_mixing_weights}
\end{equation}

\begin{equation}
    \begin{aligned}
        f_1 &= \mathcal{N}^+(50\,000, 3\,000^2) \\ 
        f_2 &= \text{Gamma}(6, 40\,000) 
    \end{aligned}
    \label{eq:c_fail_components}
\end{equation}

\begin{figure}
    \centering
    \includegraphics[width=\textwidth]{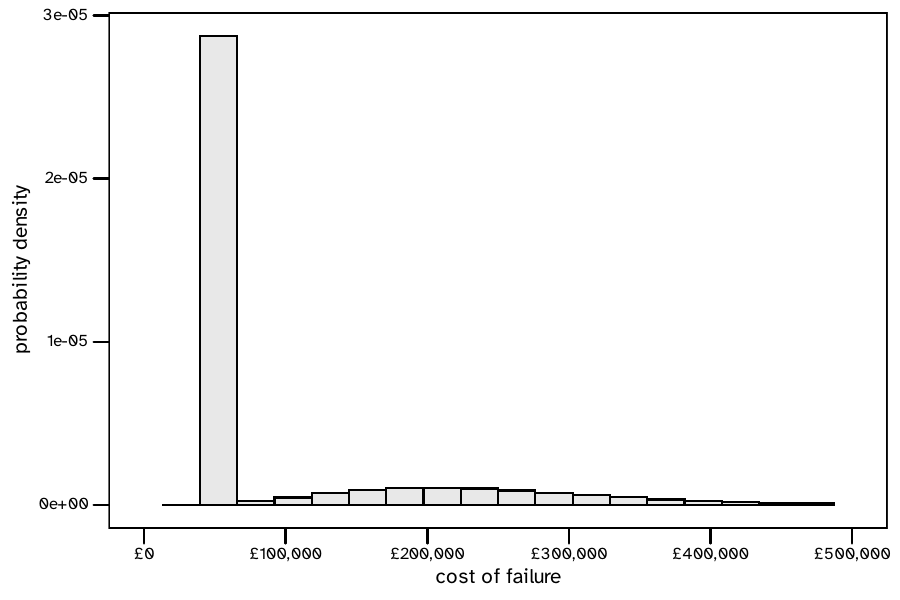}
    \caption{Histogram of samples from $\mathcal{C}_{fail}$, defined using the mixture model in Equation \ref{eq:c_fail_general} to account for uncertainties and multiple failure modes}
    \label{fig:c_fail}
\end{figure}

This represents two failure modes: minor failures (for instance leaks from pressure vessels) with lower consequences occurring approximately 75\% of the time, and major failures (such as ruptures) with potentially severe consequences occurring approximately 25\% of the time. Samples from this distribution are plotted in Figure \ref{fig:c_fail}. Note that the broader analysis presented in this paper is equally compatible with alternative cost models, including those derived from formal consequence assessment methodologies.

\subsubsection{Probabilistic Analysis of Test Set Performance} \label{sect:reliability}

Section \ref{sect:decisions} identified the need to quantify model reliability, specifically $\Pr(m_o \mid s)$ in Equation \ref{eq:exp_cost_scenario}. Traditional point estimates of classification performance fail to capture uncertainty. We therefore employ Bayesian analysis to quantify both aleatoric and epistemic uncertainty.

We model the confusion matrix rows as draws from multinomial distributions with Dirichlet priors:

\begin{equation}
    \mathbf{C}_{i,:} \sim \text{Multinomial}(N_i, \boldsymbol{\theta}_i) \quad \text{for each class } i \in \{1, 2, 3, 4\}
    \label{eq:Multinom}
\end{equation}

\begin{equation}
    \boldsymbol{\theta}_i \sim \text{Dirichlet}(\boldsymbol{\alpha}), \quad \boldsymbol{\alpha} = [1, 1, 1, 1]
    \label{eq:beta_dirichlet}
\end{equation}

where:
\begin{itemize}
    \item $\boldsymbol{\theta}_i$ represents the probability vector for classifying true class $i$ instances
    \item $\mathbf{C}_{i,:}$ denotes row $i$ of the confusion matrix
    \item $N_i = \sum_{j=1}^{4} C_{ij}$ is the total count for true class $i$
\end{itemize}

The uniform Dirichlet prior represents no preference among classes. This conjugate model yields an analytical posterior:

\begin{equation}
    \boldsymbol{\theta}_i \mid \mathbf{C} \sim \text{Dirichlet}(\boldsymbol{\alpha} + \mathbf{C}_{i,:})
    \label{eq:conj_dirichlet}
\end{equation}

This posterior distribution fully characterises classification uncertainty, explicitly representing epistemic uncertainty that decreases with additional testing. Table \ref{tab:confusion_matrix} shows the confusion matrix from 246 test radiographs.

\begin{table}[htbp]
    \centering
    \begin{tabular}{l|cccc}
     & \multicolumn{4}{c}{\textbf{Predicted Class}} \\
    \cmidrule{2-5}
    \textbf{True Class} & No anomaly & Cracking & Porosity & Lack of penetration \\
    \midrule
    No anomaly & 72 & 1 & 4 & 0 \\
    Cracking & 2 & 62 & 0 & 0 \\
    Porosity & 7 & 0 & 37 & 1 \\
    Lack of penetration & 0 & 0 & 0 & 60 \\
    \bottomrule
    \end{tabular}
    \caption{Confusion matrix for weld radiograph classification model}
    \label{tab:confusion_matrix}
\end{table}

\begin{figure}
    \centering
    \includegraphics[width=\textwidth]{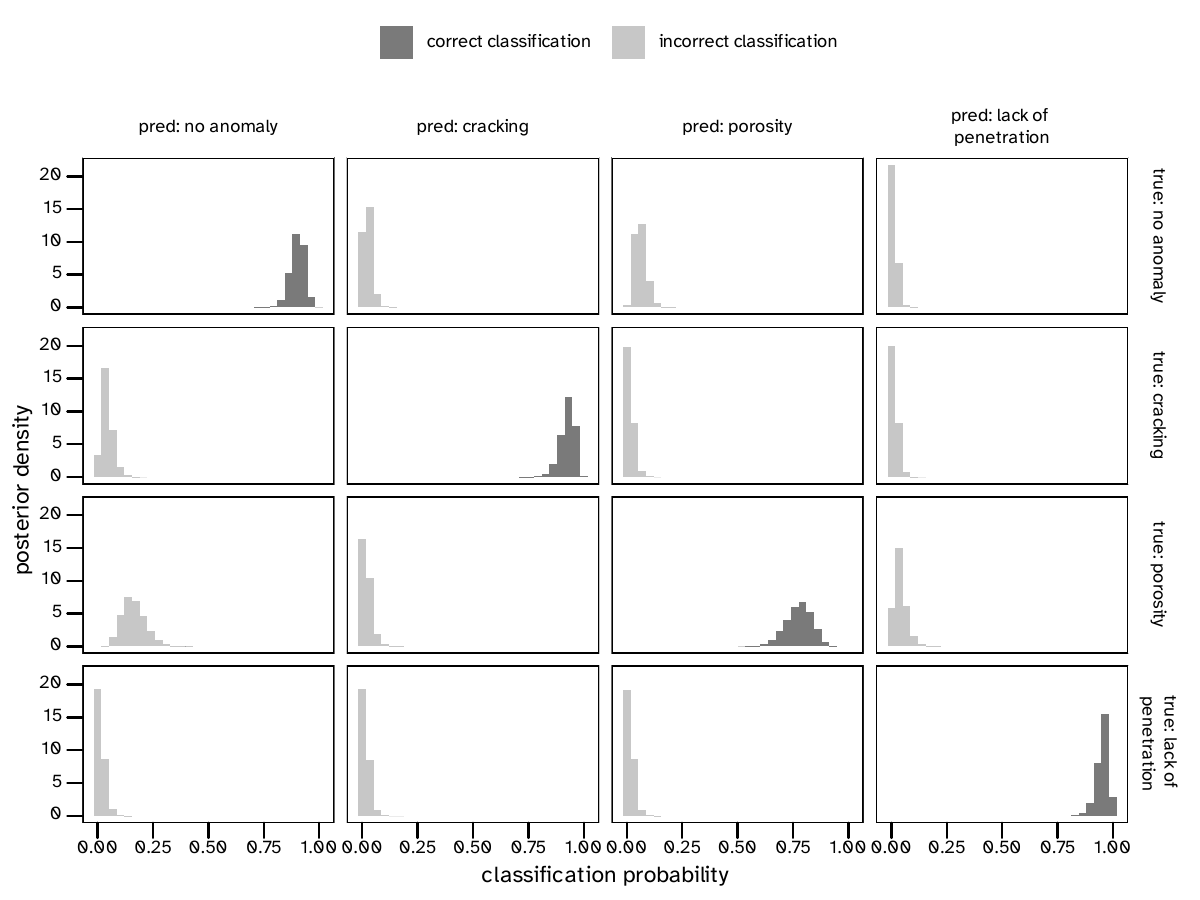}
    \caption{Marginal posterior densities for classification probabilities given true damage states}
    \label{fig:reliability_posterior}
\end{figure}

Figure \ref{fig:reliability_posterior} displays the marginal posterior densities for $\Pr(m_o \mid s)$, providing the probabilistic inputs needed for the decision analysis in Section \ref{sect:decisions}.

\begin{figure}
    \centering
    \includegraphics[width=\textwidth]{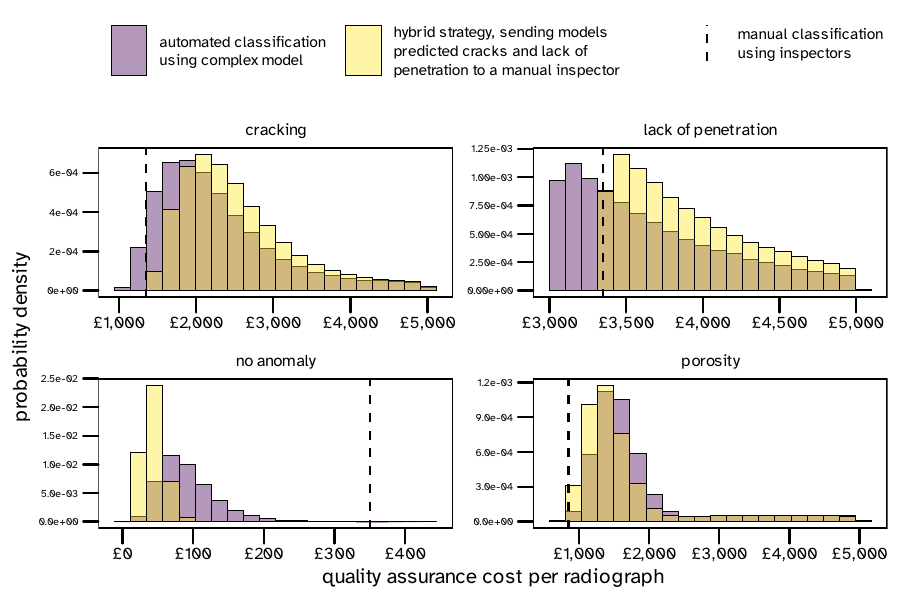}
    \caption{Expected costs by classification method and true damage state}
    \label{fig:decision_analysis}
\end{figure}

These posterior samples enable completion of the decision analysis. Figure \ref{fig:decision_analysis} presents the expected costs for three evaluation approaches across all damage scenarios.

\subsubsection{Rank-ordering of competing models}

The results in Figure \ref{fig:decision_analysis} are summarised in Table \ref{tab:decision_analysis_summary}. This shows how model risk, quantified as an expected cost, can be used to select amongst alternative approaches. Here, the results are grouped by scenario, and since the expected optimal model, $m^{*}$ differs, for a new commissioning project the decision will also require a model of anomaly density i.e. if $1,000$ welds are to be inspected, how many will contain each type of anomaly? For instance, if all actual defects were only porosity, the hybrid strategy becomes optimal if over 82.4\% of all welds have no anomaly. However, if all actual defects were cracking, over 87.2\% of welds would need to have no anomaly for the hybrid strategy to be preferable overall. Assuming an equal distribution among the three defect types, this threshold is approximately 84.5\%. 

When no damage is present, manual evaluation incurs high fixed costs, while the automated approach generates unnecessary repair costs from false positives. The hybrid strategy mitigates these automated errors through selective manual verification, achieving the lowest cost for undamaged welds. However, when damage is present, manual evaluation becomes risk-optimal across all defect types due to the severe financial consequences of the model misclassifying defects as 'no anomaly'. This risk then outweighs the manual inspection's higher baseline cost. While these results assume perfect manual inspection performance, the approach is equally compatible with realistic inspector error rates.

\begin{table}
    \centering
    \begin{tabular}{lrrr} 
    \toprule
    \textbf{true state} & \textbf{manual evaluation} & \textbf{fully automated} & \textbf{hybrid} \\
    \textbf{(scenario)} & \textbf{(\pounds{} per radiograph)} & \textbf{(\pounds{} per radiograph)} & \textbf{(\pounds{} per radiograph)} \\
    \bottomrule
    No Anomaly          & 350          & 92.55           & \textbf{43.82}  \\
    Cracking            & \textbf{1350} & 3155.79          & 3440.96          \\
    Porosity            & \textbf{850} & 2424.17          & 2282.77          \\
    Lack of Penetration & \textbf{3350} & 4501.77          & 4825.18          \\
    \bottomrule
    \end{tabular}
    \caption{Expected quality assurance and repair cost per radiograph for each model. The risk-optimal strategy ($m^*$), for each scenario, is highlighted in bold}
    \label{tab:decision_analysis_summary}
\end{table}

\subsection{Risk mitigation} \label{sect:risk_mitigation}

\subsubsection{Explainability} \label{sect:expl}

Explainable AI methods enable interrogation of complex models to understand failure modes. Two key approaches: counterfactual analysis and saliency mapping, provide complementary insights into model behaviour and decision boundaries.

Given model $m$, input $z$, and prediction $y = m(z)$, a counterfactual $z_{cf}$ satisfies $m(z_{cf}) = y_{cf}$ for a target output $y_{cf} \neq y$. We find $z_{cf}$ through gradient-based optimisation as shown in Equation \ref{eq:cf}, where $\eta$ is the learning rate and $\mathcal{L}$ is the loss function.

\begin{equation}
    z_{i+1} = z_i - \eta \times \frac{\partial}{\partial z_i} \mathcal{L} \left( m(z_i), y_{cf} \right)
    \label{eq:cf}
\end{equation}

\begin{figure}
    \centering
    \includegraphics[width = \textwidth]{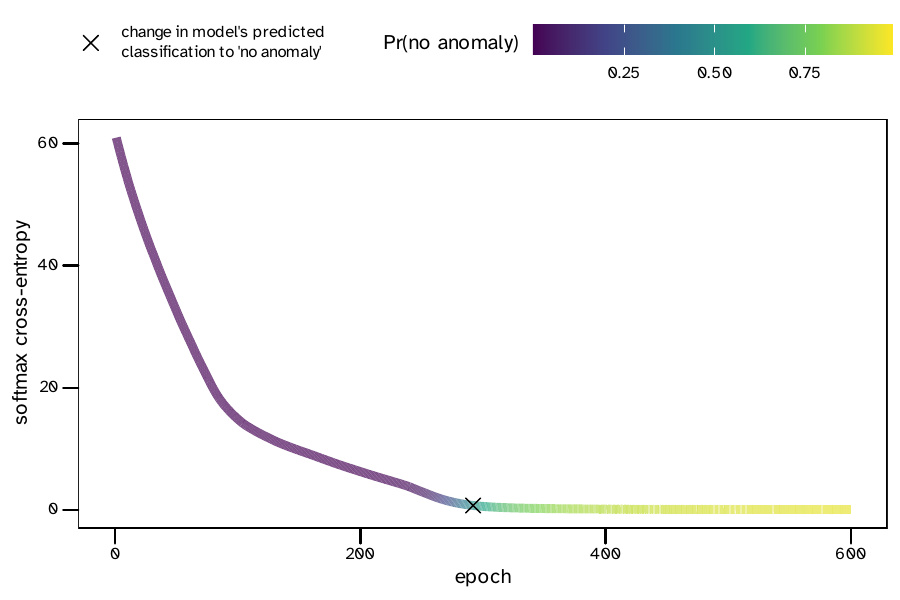}
    \caption{Loss reduction during counterfactual analysis incrementing data from a correctly classified radiograph displaying lack of penetration, to 'no anomaly'}
    \label{fig:cf_loss}
\end{figure}

\begin{figure}
    \centering
    \includegraphics[width = \textwidth]{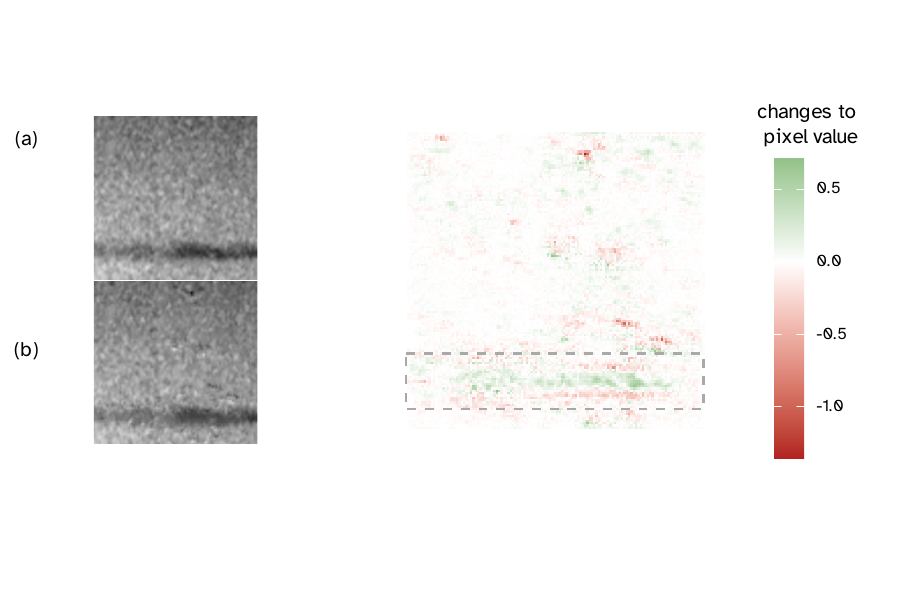}
    \caption{(a) Original radiograph image, correctly classified by trained model as "lack of penetration", and (b) The same image adjusted by a counterfactual analysis, now classified by the same model as "no anomaly"}
    \label{fig:cf_example}
\end{figure}

Figure \ref{fig:cf_loss} shows the optimisation process for transforming a radiograph correctly classified as "lack of penetration" into one classified as "no anomaly". Figure \ref{fig:cf_example} displays the original image (a) and counterfactual result (b), highlighting the pixel changes required. The dashed grey box highlights where adjustments were made that align with the visible dark regions (missing weld metal) on the radiographs.

Saliency maps quantify each input pixel's influence on model predictions. For class $c$, the saliency $S_{i,j}$ of pixel $z_{i,j}$ is shown in Equation \ref{eq:saliency}, where $S_c(z) = \Pr(y = c \mid z)$ is the class score.

\begin{equation}
    S_{i, j} = \left| \dfrac{\partial S_{c}(z)} {\partial z_{i, j}} \right|
    \label{eq:saliency}
\end{equation}

\begin{figure}
    \centering
    \includegraphics[width = \textwidth]{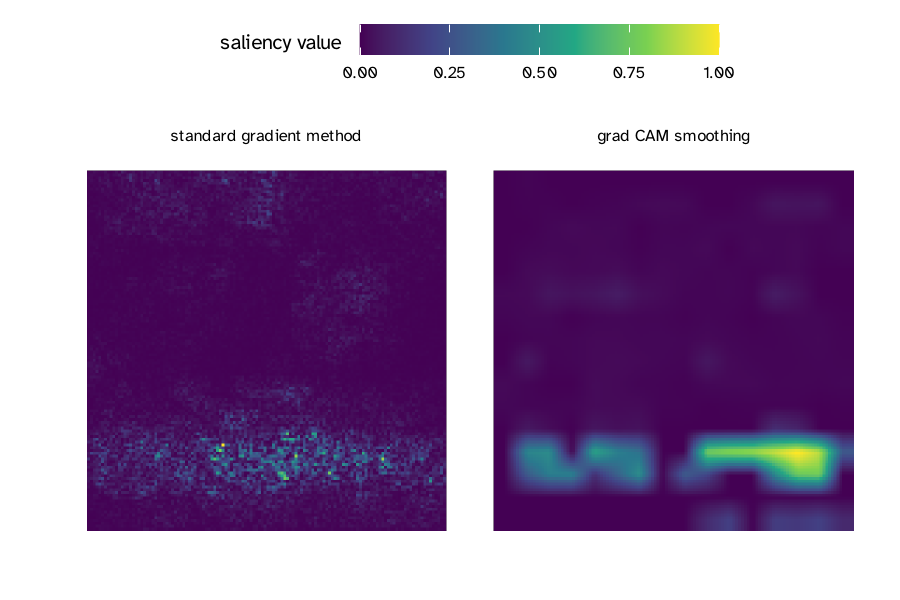}
    \caption{Saliency map of pixel importance for original radiograph image, correctly classified by trained model as a lack of penetration. Higher saliency values indicate pixels that were more consequential in determining the outcome classification.}
    \label{fig:saliency}
\end{figure}

Figure \ref{fig:saliency} presents saliency maps for the radiograph from Figure \ref{fig:cf_example}(a). Both standard saliency and GRADient-weighted Class Activation Mapping (Grad-CAM)\footnote{Grad-CAM uses gradients from the final convolutional layer for noise reduction.} highlight the dark horizontal line indicating missing weld metal. This alignment with the visible defect is evidence towards the model's valid decision process. Misalignment between high-saliency regions and engineering expectations would signal either novel feature detection or poor generalisation, both requiring risk assessment.

\subsubsection{Model verification} \label{sect:voi}

Higher risk applications will require at least as much verification (and generally more) when compared to comparatively low risk applications. One approach for providing a quantitative rationale for identifying a proportionate extent of verification is Value of Information (VoI) analysis \cite{Raiffa1966}. 

In the context of model verification, VoI analysis compares two scenarios:
\begin{enumerate}
    \item \textbf{Prior decision}: Optimised costs conditional on current uncertainty estimates, see Equation \ref{eq:prior_costs}.
    \item \textbf{Preposterior decision}\footnote{a term popularised in engineering applications of VoI analysis \cite{Jordaan2005}}: Optimised costs after updating models with prospective new verification data, see Equation \ref{eq:prepost_costs}.
\end{enumerate}

Value of Perfect Information (VoPI) represents the special case where $z$ eliminates all uncertainty in model parameter(s), $\theta$. While unrealistic, VoPI is simpler to compute and provides an upper bound on VoI (which asymptotically approaches VoPI as data quality increases \cite{DiFrancesco2023}).

\begin{equation}
    \mathcal{C}_{\text{Prior}} = \min_{m \in M} \mathop{\mathbb{E}}_{\Pr(\theta)} [\mathcal{R}_m(m, \Pr(m_o|\theta, s))]
    \label{eq:prior_costs}
\end{equation}

\begin{equation}
    \mathcal{C}_{\text{Pre-posterior}} = \mathop{\mathbb{E}}_{\Pr(z)} \left[ \min_{m' \in M} \mathop{\mathbb{E}}_{\Pr(\theta|z)} [\mathcal{R}_{m}(m', \Pr(m_o|\theta, s))] \right]
    \label{eq:prepost_costs}
\end{equation}

\begin{equation}
    VoI(s) = \mathcal{C}_{\text{Prior}} - \mathcal{C}_{\text{Pre-posterior}}
    \label{eq:vopi}
\end{equation}

\begin{algorithm}
    \caption{Value of Perfect Information for Model Verification}
    \label{alg:vopi_verification}
    \begin{algorithmic}[1]  % <-- YOU NEED THIS
        \ForAll{scenarios $s \in S$}
            \ForAll{models $m \in M$}
                \ForAll{samples from probabilistic model of reliability, $\theta_{i} \sim \Pr(\theta)$}
                    \State calculate associated model risk, using Equation \ref{eq:exp_cost_scenario}
                \EndFor
            \EndFor
            \State \textbf{Prior Analysis:}
            \State identify prior optimal costs, using Equation \ref{eq:prior_costs}
            \State \textbf{Pre-posterior Analysis:}
            \State identify pre-posterior optimal costs, using Equation \ref{eq:prepost_costs}
            \State calculate VoPI using Equation \ref{eq:vopi}
        \EndFor
        \State weight outcomes by $\Pr(s)$ to obtain expected model risk and value of information averaged over all scenarios.
    \end{algorithmic} % <-- AND THIS
\end{algorithm}

\begin{figure}
    \centering
    \includegraphics[width=\textwidth]{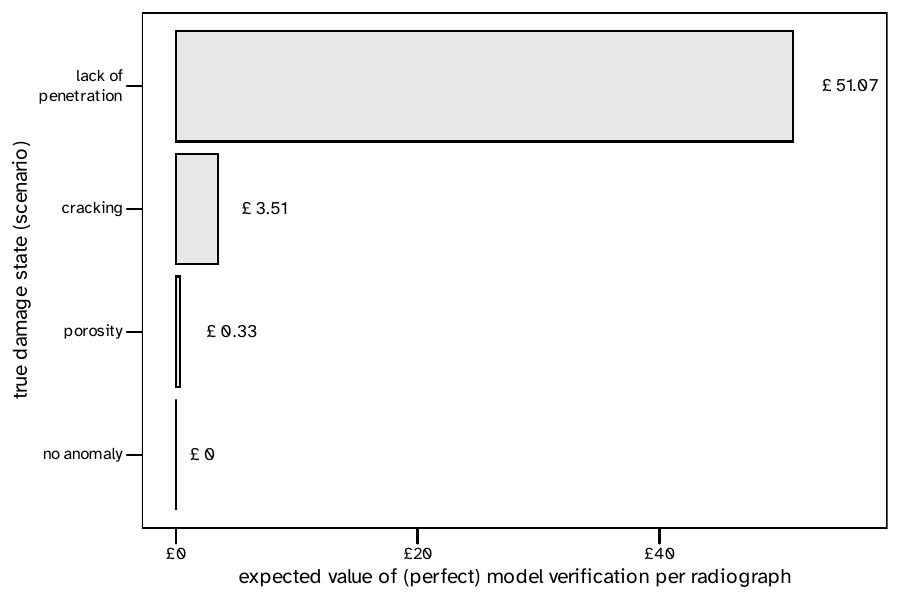}
    \caption{Expected value of (perfect) model verification for each scenario considered}
    \label{fig:vopi}
\end{figure}

Figure \ref{fig:vopi} shows the expected value of perfect verification for each damage scenario. Lack of penetration exhibits the highest value because the expected costs of different evaluation strategies are closely matched (see Figure \ref{fig:decision_analysis}), making the optimal choice highly sensitive to model reliability. In such cases, even small reductions in uncertainty of model performance can shift the decision boundary, justifying investment in verification. Conversely, the "no anomaly" scenario is not expected to benefit from further verification. This is not because the model is perfect, but because an improved understanding of model reliability is not expected to identify a new optimal strategy or reduce exposure risk.

For a project with $100$ radiographs where a lack of penetration was present, further model verification would be worth up to \pounds $5,107$. If verification costs are quoted to be above this threshold, then further testing would not be considered a risk-optimal investment.

This demonstrates a key principle: verification resources should be concentrated where the information will most benefit decision-making, and not necessarily where model uncertainty is greatest. By quantifying the economic value of reduced uncertainty in this way, it is possible to identify when model risk has reached ALARP i.e. the point where further verification costs exceed their benefits to reduce risk.

\begin{comment}

\subsection{Probabilistic modelling of corrosion growth rate} \label{sect:examples_bayes}

\textcolor{red}{incomplete}

The development of probabilistic programming languages cite[Stan, Turing, NumPyro] has democratised scalable uncertainty quantification.

%\subsection{Asset Management of Fleets} \label{sect:examples_rl}

\textcolor{red}{incomplete}
    
Forecast maintenance for interventions on a fleet of assets. When attempting to define this algorithmically, we see that the system can exist in many (in principle, an infinite number of) states.

This is a typical use case for reinforcement learning solutions. Here is what a reinforcement learning solution might look like:

learn a policy from a reward signal, obtained from running simulations. Maximising this reward is the aim of the game, and so it is assumed that it describes all desirable (and suitably penalises undesirable) behaviour from a system - sometimes referred to as the \emph{reward hypothesis}. The policy describes how to act given a current state and the (unknowable) solution we are trying to approximate is the policy that maximises the total future reward. Accounting for possible delayed consequences, makes this challenge large, and is why simulation methods are sometimes adopted. It will cost me money to schedule repairs now, but will it decrease downtime overall for the remainder of the asset life? Or will

Is following this policy safe?

\end{comment}

\section{Conclusions} \label{sect:conclusions}

Complex models offer significant potential benefits but introduce an associated model risk (the expected consequence of incorrect outputs). In this paper the ALARP principle is proposed to guide safety-critical industries in evaluating whether model risk has been reduced to acceptable levels. While illustrated through automated weld radiograph classification, the framework is considered to apply broadly to any domain where computational models inform high-consequence decisions.

Safety auditors should independently evaluate model risk through the following steps:

\begin{enumerate}
    \item \textbf{identify model-decision boundaries}:
    duty holders should document how model outputs translate into operational decisions and their potential consequences, with respect to the model's precise operational domain, credible operational scenarios, potential model failure modes and their downstream consequences.
    
    Section \ref{sect:risk_quant} demonstrated this using decision diagrams to map the relationships between model predictions, interventions, and costs. This step is critical for any complex model—from predictive maintenance algorithms to clinical decision support systems—as risk only materializes when models inform actions. Methods from explainable/adversarial AI were introduced in Section \ref{sect:expl} as approaches that could uncover failure modes.

    \item \textbf{review model risk management strategy}: duty holders should have completed a review to find sources of model risk. Model performance should then be quantified probabilistically, as "cherry picked" instances, and point estimates are not sufficiently informative to evaluate risk.

    Section \ref{sect:risk_quant} presets a Bayesian analysis of test set results from a classification model, to obtain a probabilistic estimate of model reliability that is compatible with decision analysis.

    \item \textbf{apply proportionate verification}
    The extent of risk mitigation should reflect the consequence level of the application. The safety case should be reviewed in response to monitoring outcomes, model retraining, or operational changes that invalidate any underlying assumptions of the risk assessment.

    In this example, model reliability was quantified using a statistical analysis of a test set evaluation and a value of information analysis was used to quantify how much an operator should be willing to pay for further verification testing in Section \ref{sect:risk_mitigation}.
    
    \item \textbf {governance, roles, and responsibilities}: a governance structure is required to oversee the use of complex models, with responsibilities for model development, independent validation, monitoring, and risk management. Such a framework has reached mature development in banking today, where both regulatory pressures and internal organisational systems now exist. 
    
    This was not explicitly demonstrated in the example calculations in this paper, but the literature review highlighted it as an important component of model risk management.

    \item \textbf{document ALARP determination}: when a risk-optimal modelling strategy has been identified, compile evidence that the associated residual model risk is tolerable and further reduction would be disproportionate. 
    
    Sections \ref{sect:risk_quant} and \ref{sect:risk_mitigation} provides a template for this documentation, showing how quantitative risk assessment, probabilistic reliability analysis, and economic evaluation combine to support ALARP arguments.
    
\end{enumerate}

 The above are considered to align with existing high level government, industrial and academic guidance, as well as the detailed example of quantifying model risk for automating weld radiograph classification presented in this paper.

\section{Acknowledgements}

Domenic Di Francesco is supported by the Ecosystem Leadership Award under the EPSRC Grant EP/X03870X/1, and The Alan Turing Institute, particularly the Turing Research Fellowship scheme under that grant.

Some computations described in this paper were performed using the Baskerville Tier 2 HPC service (https://www.baskerville.ac.uk/). Baskerville was funded by the EPSRC and UKRI through the World Class Labs scheme (EP/T022221/1) and the Digital Research Infrastructure programme (EP/W032244/1) and is operated by Advanced Research Computing at the University of Birmingham.

The authors would like to thank our colleagues for reviewing the work and providing useful inputs, in particular Graeme West (University of Strathclyde, The Alan Turing Institute) and Justin Bunker (University of Cambridge).

\bibliography{refs}

\end{document}